comp-gas/9403006 31 Mar 94

# APPLICATION OF COMPLEX DAUBECHIES' WAVELETS TO NUMERICAL SIMULATION OF A NONLINEAR SIGNAL PROPAGATION MODEL*

*L. Gagnon*[1,2], *J. M. Lina*[1,3] and *B. Goulard*[1]

[1]Labo. de Phys. Nucl., Univ. de Montréal, C.P. 6128, Succ. Centre-Ville, Montréal, Québec, H3C 3J7, Canada

[2]Collège Militaire Royal de St.-Jean, St-Jean-sur-le-Richelieu, Québec, J0J 1R0, Canada

[3]Atlantic Nuclear Services Ltd., Fredericton, New Brunswick, E3B 5C8, Canada

e-mail: lgagnon@lps.umontreal.ca

## Abstract

We report the first application of complex symmetric wavelets to the numerical simulation of a nonlinear signal propagation model. This model is the so-called nonlinear Schrödinger equation that describes, for instance, the evolution of the electric field amplitude in nonlinear optical fibers. We propose and study a new way to implement a global space-time adaptive grid, based on interpolation properties of higher-order scaling functions.

## 1. Introduction

The idea of using wavelets to perform numerical simulations of partial differential equations is not new (see, for instance, [1, 2] and references therein). The motivations come from the fact that wavelets provide a mathematical representation that can resolve numerical difficulties due to singular phenomena. More exactly, wavelet properties such as orthogonality and compact support, as well as exact representation of polynomials of a fixed degree in wavelet bases, allow efficient and stable calculation of regions with transient phenomena or strong oscillations. Up to now, wavelets that have been used for such studies are the "classical" ones (real Daubechies wavelets [3], splines, Shannon and Meyer wavelets, etc.).

In the present work, we are interested in using complex symmetric wavelets [4] as an interpolating tool in the numerical sampling process. To this aim, we will consider the nonlinear signal propagation model

$$iu_t + \frac{1}{2}u_{xx} + \lambda u|u|^2 = 0; \quad \lambda \in \mathbb{R}^*, \tag{1.1}$$

that is known as the nonlinear Schrödinger (NLS) equation. Equation (1.1) describes, for instance, the evolution of the electric field amplitude $u(t,x)$ in nonlinear optical fibers (in this particular application, $t$ and $x$ are space and time variables respectively). In this paper, we use the notation NLS"+" for $\lambda > 0$ and NLS"−" for $\lambda < 0$.

*Work supported in part by the Natural Sciences and Engineering Research Council (NSERC) of Canada

The choice of the model (1.1) is interesting because it can lead to strong gradient phenomena. For the NLS"−" case, for instance, smooth localized initial conditions can evolve toward a typical breaking wave phenomena that exhibits very high frequency oscillations [5, 6]. In addition, breather solutions (also known as the bound N-soliton solutions in optics) of the NLS"+" equation, shows periodic peaking of the signal amplitude (see, for instance, [7] and references therein) and can be used to test the ability of an adaptive algorithm to increase *and* decrease the grid.

The article is organized as follows. In Section 2, we recall the discrete wavelet theory through the multiresolution analysis and give an example of complex symmetric wavelets. In Section 3, we study a simple way to implement a global space-time adaptive discretisation in the split-step Fourier method based on the recomposition properties of higher-order scaling functions. This globally adaptive algorithm will be tested on typical simulations (optical breaking waves and N-soliton solutions).

## 2. Complex symmetric wavelets

Wavelets can easily be introduced using the multiresolution analysis [8].

The basic equation of the multiresolution theory is the scaling equation that establishes a link between the underlying symmetries of the wavelet theory: dilations and translations. Given a set of complex coefficients $a_k, k \in \mathbb{Z}$, the scaling equation

$$\varphi(x) = 2\sum_k a_k \varphi(2x-k), \quad x \in \mathbb{R} \tag{2.1}$$

with the normalization $\int \varphi(x)\, dx = \sum_k a_k = 1$, define a scaling function $\varphi(x)$.

The multiresolution analysis is a particular decomposition of the space of square integrable functions $L^2(\mathbb{R})$ into a chain of closed subspaces $\ldots V_{j-1} \subset V_j \subset V_{j+1} \subset \ldots, j \in \mathbb{Z}$, where the set $\{\varphi_{j,k}(x) = 2^{\frac{j}{2}}\varphi(2^j x - k), k \in \mathbb{Z}\}$ is a basis for $V_j$. The index $j$ represents the scale to which the basic scaling function $\varphi(x)$ is dilated. For a given scale, the space $V_j$ is generated by the translations of $\varphi_{j,0}(x)$ by the amount $k2^{-j}$.

Multiresolution also allows to decompose $L^2(\mathbb{R})$ as

$$L^2(\mathbb{R}) = V_{N_0} \oplus \sum_{j \geq N_0} W_j, \qquad (2.2)$$

where $W_j$ is the orthogonal complement of $V_j$ in $V_{j+1}$ ($V_{j+1} = V_j \oplus W_j$) and $N_0$ is the coarsest resolution level.

For a given scale $j$, $W_j$ is generated by the set $\{\psi_{j,k}(x) = 2^{\frac{j}{2}}\psi(2^j x - k), k \in \mathbb{Z}\}$, where the analysing wavelet $\psi(x)$ satisfies

$$\psi(x) = 2\sum_k b_k \varphi(2x - k), \quad x \in \mathbb{R}, \qquad (2.3)$$

with $b_k = (-1)^k \bar{a}_{1-k}$. Therefore, any function $f(x)$ of $L^2(\mathbb{R})$ can be expanded as a linear combination of translates of the scaling equation at the scale $N_0$ and the translates of the associated wavelet at the smaller scales $N_0 + 1, N_0 + 2, ...$, as

$$f(x) = \sum_k v_{N_0,k}\varphi_{N_0,k}(x) + \sum_{j \geq N_0} \sum_k w_{j,k}\psi_{j,k}(x), \quad (2.4)$$

where $w_{j,k}$ are the wavelet coefficients (that are non-negligible only in high gradient regions). In practice, expansion (2.6) is computed by fixing a maximal resolution $N$ for which the wavelet coefficients $w_{N,k}$ are negligible, that is,

$$f(x) = \sum_k v_{N,k}\varphi_{N,k}(x). \qquad (2.5)$$

In this work, we will consider Daubechies' wavelets and, in particular, the recently investigated complex symmetric cases [4]. Important numerical features of wavelets (compact support, orthogonality and regularity) can be implemented in the multiresolution analysis as the following. Compact support requires a finite number of scaling coefficients $a_k$ in (2.1). Orthonormality condition imposes this number to be even. In this work we choose $k = -J, -J + 1, ..., J + 1$, which makes $\varphi(x)$ and $\psi(x)$ non-zero inside the interval $[-J, J + 1]$. Symmetry of the scaling function $\varphi(x)$ with respect to $x = 1/2$ is possible only for complex scaling coefficients. Due to the localisation properties of $\varphi_{N,k}(x)$ around the sampling points $x_{N,k} = \frac{1+2k}{2^{N+1}}$, Taylor expansion of $f(x)$ in $v_{N,k} = \int \bar{\varphi}_{N,k}(x)f(x)dx$ then leads to

$$v_{N,k} \cong 2^{-N/2} f(x_{N,k}). \qquad (2.6)$$

Coefficients $v$ and $w$ at lower resolutions are obtained from equations (2.1) and (2.3). Finally, regularity amounts to require that the first polynomial terms of the Taylor expansion of a function $f$ are mapped exactly in $V_j$. For a given order $J$, we can require such condition up to polynomials of degree $J$. An example of complex symmetric scaling functions and wavelets that result from the above requirements is given in Figure 1 for $J = 4$.

In the following section, we will use this wavelet to detect strong gradients in the signal $u(t, x)$ and to dynamically change the sampling rate.

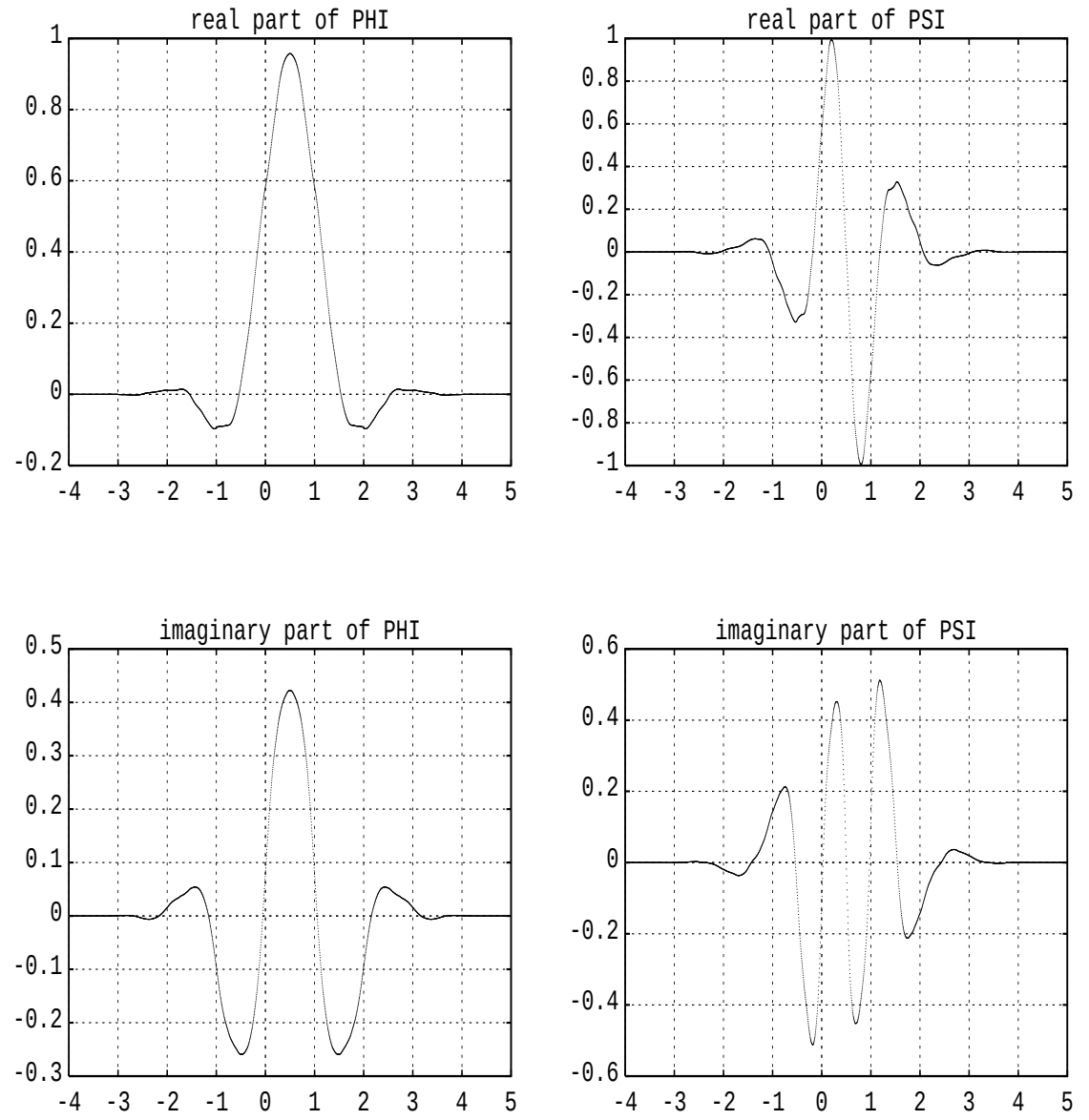

Figure 1: Real and imaginary parts of $\varphi(x)$ and $\psi(x)$ for the complex symmetric case $J = 4$

## 3. A globally adaptive algorithm

One of the most popular numerical scheme to solve NLS-type equations is the so-called "symmetrized split-step Fourier method" (see, for instance, [7] and references therein). This is essentially a pseudo-spectral algorithm that splits the linear and nonlinear effects over one time step. Because nonlinear terms are more easily calculated in the physical space, only the linear part is decomposed in its Fourier components using a FFT algorithm.

We now want to demonstrate how regularity properties of scaling functions can be used to increase the level sampling $N$ over the entire x-window during the simulation. In particular, this will result in an efficient hybrid numerical algorithm that combines the rapidity of the FFT for calculating the time evolution, and the interpolation properties of the scaling functions for changing the sampling rate.

The idea is the following. At the beginning of each time step, one performs a one level wavelet transform of the signal $u(t, x) \in V_N$ from the sampling level $N$ to level $N - 1$, i.e. $u(t, x) = \sum_k v_{N-1,k}\varphi_{N-1,k}(x) + \sum_k w_{N-1,k}\psi_{N-1,k}(x)$, in order to detect the presence of strong gradients. When the maximal absolute value of the wavelet coefficients $w_{N-1}$ is greater that a predetermined threshold (fixed to 0.005 in our simulations), the num-ber of collocation points is doubled using an interpolation process. This is done by performing an inverse wavelet transform, from level $N$ to level $N+1$, on the coefficients $v_N$. To this aim, the $2^N$ wavelet coefficients $w_N$ are set to zero and the result of the inverse transformation is a set of $2^{N+1}$ scaling coefficients $v_{N+1}$ that constitute the new signal sampling (after multiplying by a $\sqrt{2}$ normalization factor).

In the above scheme, an important point to be careful about is that no local asymmetry is introduced during the resampling process. This is where symmetry properties of the scaling functions presented in Section 2 are useful. Unfortunately, a recomposition using these complex scaling functions, slightly couples the real and imaginary parts of the signal $u(t,x)$ (for instance, by introducing imaginary components in $v_{N+1}$ when the original signal is real). This can be avoided, by resampling the signal, using a linear combination of the inverse wavelet transform and its complex conjugate; in other word, by recomposing the signal using the real part of the coefficients $a_k$ only. In fact, we have found this recomposition scheme very accurate when done with higher-order wavelets ($J \geq 8$).

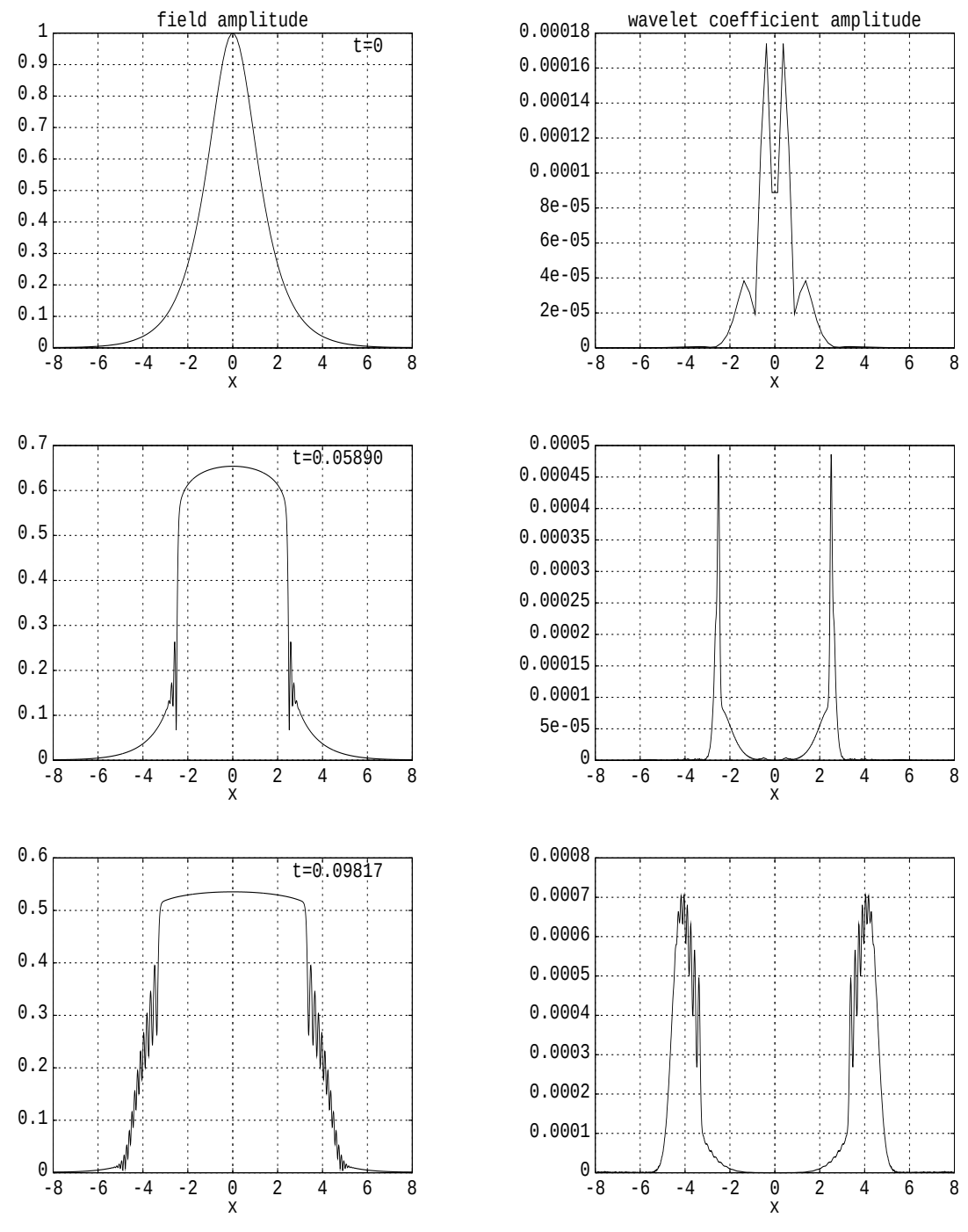

Figure 2: Field and wavelet coefficient amplitude for a typical breaking wave simulation

We have written a C-program that simulates (1.1) using the split-step Fourier method and the global resampling scheme described above [9]. In order to conserve the numerical stability, we have also chosen to decrease the time step by a factor of 4 after each resampling. The maximum wavelet amplitude for the resampling threshold has been fixed to 0.005.

Figure 2 shows a first simulation that leads to strong gradients. This is a typical breaking wave phenomena that appears in nonlinear optical fibers in the normal dispersion regime, that is for the NLS"−" equation with $\lambda < 0$. In this simulation, $\lambda = -900$, the initial signal is $u(0,x) = sech(x)$ and has been sampled 128 times, the initial time range is $\pi/32$ and the initial time step has been set to $\pi/1600$. The amplitude of the wavelet coefficients $w_{N-1}(t)$ is quite enlightning the sharp signal structure evolution. Figure 3 shows the evolution of the maximal amplitude of these wavelet coefficients as $t$ increases. Since each peak is the result of a new sampling, we end up with a final number of 2048 collocation points. A comparison with a non-adaptive simulation (with 2048 samples and the corresponding $\Delta t = \pi/409600$) has shown that the accuracy of the adaptive simulation is quite satisfactory. In fact, no difference can be detected on the amplitude and phase from the plots.

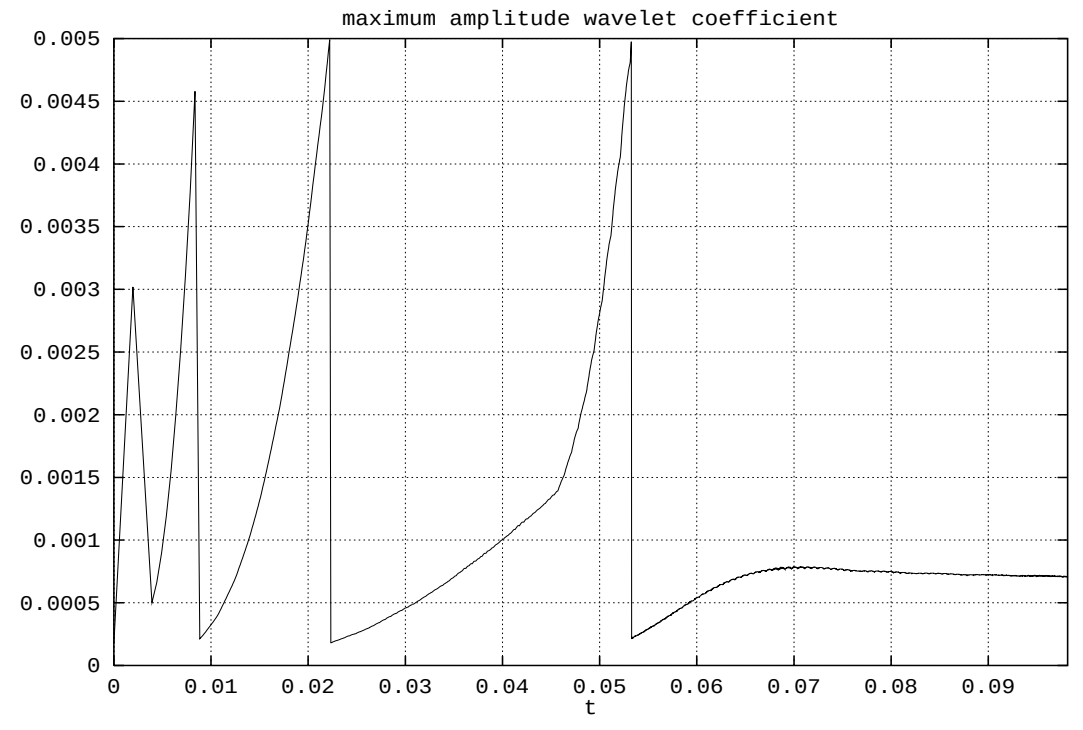

Figure 3: Variation of the maximum amplitude wavelet coefficient for the simulation of Figure 2

The above adaptive algorithm can also be modified to allow a decreasing in the number of sampling points when sharp gradients tend to disappear. Similarly as above, the new signal sampling can be obtained from the scaling coefficients of a one level real-valued wavelet decomposition (after multiplying by the normalisation factor $1/\sqrt{2}$).

Interesting solutions that can be used to test the "desampling" process are the N-soliton solutions of the NLS"+" equation. These are solutions for which the amplitude evolves periodically in time, with a period of $\pi/2$, through successive amplitude peakings. They correspond to the initial conditions $u(0,x) = N sech(x)$, where $N$ is the solution order (the number of solitons involved in the solution).

Figure 4 shows the variation of the maximal amplitude of the wavelet coefficients for the case $N = 3$, over one period ($t = \pi/2$) with 128 initial samples and $\Delta t = \pi/2560$. The upper threshold has been fixed to 0.005 and the lower one to 30% of the maximal wavelet amplitude after the first resampling. This is to avoid a premature decreasing of the sampling due to possible oscillatory structures during a resampling relaxation time. The number of sampling points varies as 128-256-512-1024-512-1024-512-256-128, following the periodicity in the signal amplitude. For this particular simulation, we have measured a 10% precision on the final signal phase, with respect to the theoretical value of $\pi/4$. On the other hand, the discrepancy of the signal amplitude, with respect to the theoretical value, turns out to be very small. This is very satisfactory, if we take into account that 4 resamplings and 4 “desamplings” have been performed during the simulation.

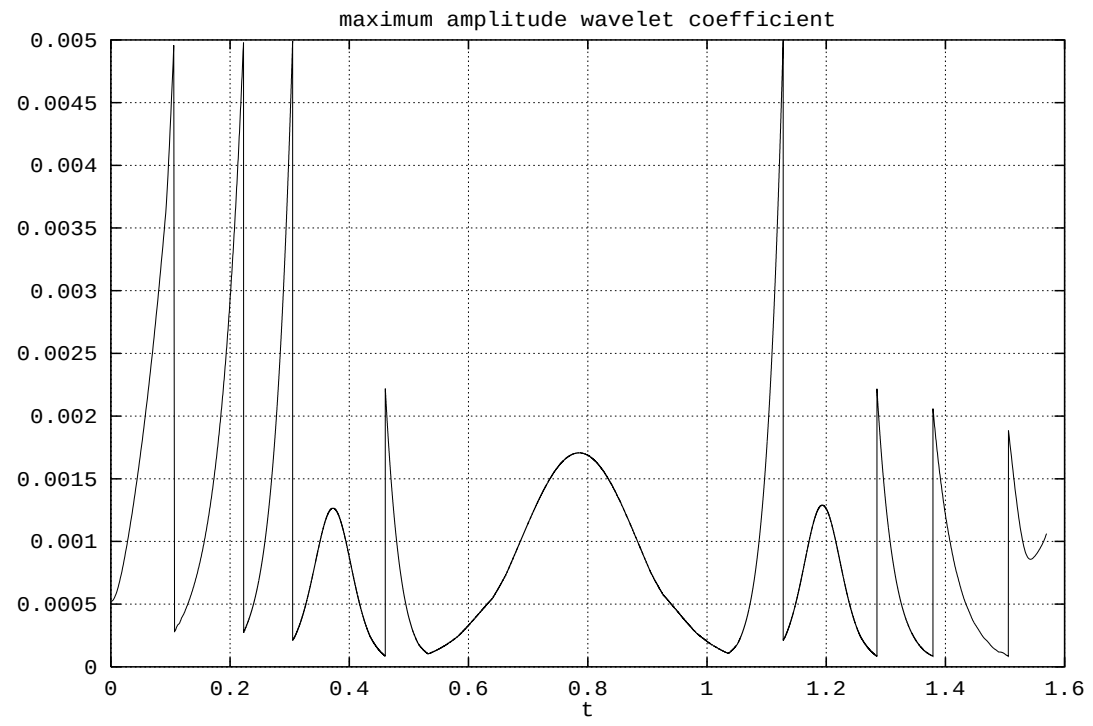

Figure 4: Variation of the maximum amplitude wavelet coefficient for the 3-soliton solution

### 4. Conclusion and future works

Wavelets have various fields of applications that go well beyond usual signal analysis [10]. Numerical simulation on wavelet bases is one of them. Here, we have proposed and studied a simple way to dynamically change the sampling rate of a signal that evolves through high gradient phenomena. The method was based on the interpolation properties of highly regular complex symmetric scaling functions. We have tested our scheme on the physically relevant nonlinear Schrödinger model for various high gradient simulations: optical breaking wave, collapse (not reported here) and bound solitons. In all these simulations, the resampling process turned out to be stable and very accurate.

An extension of the present work that is actually under investigation [9] involves the wavelet-based adaptive numerical method developped in Refs. [1, 2]. Briefly, the idea is to perform the linear evolution on the wavelet spaces rather than the Fourier one, by decomposing the signal on $\ldots W_{N-3} \oplus W_{N-2} \oplus W_{N-1}$. The main advantage is that wavelet coefficients that have amplitudes below a certain threshold (for instance, 0.0005) can be discarded in the numerical algorithm, without introducing significant numerical error. Calculating the evolution on these coefficients only, amounts to use a spatial grid that is locally adapted to the signal gradients. Our plan is to calculate the signal evolution over one time step using the same ”split-step Fourier algorithm” except that the Fourier transform is replaced by a wavelet transform, together with the appropriate representation of the second derivative operator. The advantage is that nonlinear effects can be calculated in the space $V_N$ using a simple collocation technique. In addition, we want to keep the possibility of globally resampling the signal from $V_N$ to $V_{N+1}$, using the method described in Section 3, when the maximum absolute value of $w_{N-1}$ increases above a fixed threshold (for instance, 0.005).